# Polarization due to rotational distortion in the bright star Regulus

Daniel V. Cotton[1,*], Jeremy Bailey[1], Ian D. Howarth[2], Kimberly Bott[3,4], Lucyna Kedziora-Chudczer[1], P. W. Lucas[5], J. H. Hough[5]

[1]School of Physics, UNSW Sydney, NSW 2052, Australia.

[2]Department of Physics and Astronomy, University College London, Gower Street, London, WC1E 6BT, UK.

[3]Virtual Planetary Laboratory, Seattle, WA 98195, USA.

[4]University of Washington Astronomy Department, Box 351580, UW Seattle, WA 98195, USA.

[5]Centre for Astrophysics Research, School of Physics, Astronomy and Mathematics, University of Hertfordshire, Hatfield AL10 9AB, UK.

*Correspondence to:  d.cotton@unsw.edu.au

**Polarization in stars was first predicted by Chandrasekhar[1] who calculated a substantial linear polarization at the stellar limb for a pure electron-scattering atmosphere. This polarization will average to zero when integrated over a spherical star but could be detected if the symmetry is broken, for example by the eclipse of a binary companion. Nearly 50 years ago, Harrington and Collins[2] modeled another way of breaking the symmetry and producing net polarization – the distortion of a rapidly rotating hot star. Here we report the first detection of this effect. Observations of the linear polarization of Regulus, with two different high-precision polarimeters, range from +42 parts-per-million (ppm) at a wavelength of 741 nm to –22 ppm at 395 nm. The reversal from red to blue is a distinctive feature of rotation-induced polarization. Using a new set of models for the polarization of rapidly rotating stars we find that Regulus is rotating at $96.5^{+0.6}_{-0.8}$% of its critical angular velocity for breakup, and has an inclination greater than 76.5 degrees. The rotation axis of the star is at a position angle of 79.5±0.7 degrees. The conclusions are independent of, but in good agreement with, the results of previously published interferometric observations of Regulus[3]. The accurate measurement of rotation in early-type stars is important for understanding their stellar environments[4], and course of their evolution[5].**

Initial efforts to detect Chandrasekhar's polarization effect proved unsuccessful, leading instead to the discovery of interstellar polarization[6,7]. The predicted polarization in an eclipsing binary has been observed only once, in Algol[8]. The early models of these effects used pure electron scattering atmospheres[1,6]. When more realistic, non-gray stellar-atmosphere models were applied to rotating stars[9,10] the expected polarization at visible wavelengths was found to be much smaller. Indeed, the low predicted polarizations were used to infer that this mechanism could not explain the large polarizations observed[11] in Be stars, which are now attributed to circumstellar decretion disks[12,13].

The development, over the last decade or so, of a new generation of polarimeters[14-16] that can measure stellar polarization at the ppm level brings the small atmospheric polarization effects into the range of observations. Using one of these instruments, PlanetPol[15] on the 4.2-m William Herschel Telescope, Regulus' polarization was measured as 36.7±0.8 ppm[17], which stood out from the much smaller polarizations measured for other stars at similar distance in the same part of the sky. It was concluded that the polarization was likely to be intrinsic. Since Regulus is a rapidly rotating star, polarization due to rotational distortion was considered a possibility. Sonneborn[10] had calculated models for rapidly rotating stars of types B0, B1, B2 and B5 and extrapolation of these to Regulus' B7V spectral type suggested that the polarization was about what was expected for a star rotating at 95% of the critical angular velocity for breakup; very similar to the rotation rate determined from interferometric observations at the time[5]. However, as only one wavelength was observed further observations and modeling were needed to confirm the interpretation.

Here we combine the PlanetPol observations with new observations of Regulus obtained with the High Precision Polarimetric Instrument, HIPPI[14] on the 3.9-m Anglo-Australian Telescope at Siding Spring Observatory. These observations were made in four broad wavelength bands with effective wavelengths of 395, 462, 596 and 618 nm. The PlanetPol observations are in a redder band with an effective wavelength of 741 nm. In Figure 1 we show the measured polarizations averaging all observations in each band and correcting for a small amount of interstellar polarization (see Methods). The data are plotted in the form of a QU diagram, where $Q/I$ and $U/I$ are related to the degree of polarization $p$ and position angle $\theta$ (measured eastwards from north) through $Q/I = p \cos 2\theta$ and $U/I = p \sin 2\theta$.

It can be seen from Figure 1 that the polarization varies strongly with wavelength, and the points lie nearly on a line passing through the diagram's origin. This is exactly what is expected for polarization due to rotational distortion, where gravity darkening toward the equator induces a wavelength dependent asymmetry in the stellar disk. The models of Sonneborn[10] show just this reversal of direction with the polarization being parallel to the star's rotation axis at red wavelengths, and perpendicular to the star's rotation axis in the UV (when the co-ordinate system is rotated to align Stokes Q with the star's rotation axis this corresponds to positive and negative polarization respectively, and we present data by this convention later). In Sonneborn's B5 model the reversal of polarization occurs at ~380 nm. In our Regulus observations it occurs at ~480 nm. However, this is entirely consistent with the trend shown in Sonneborn's models of the reversal shifting to longer wavelengths for later-type stars.

Several factors give us confidence in the reliability of the observations. All wavelength bands have been observed multiple times. The repeat observations agree very well even with observations separated by months to years (see Methods). There is no indication of any variability. We can exclude the possibility of Regulus' companion significantly affecting our measurements (see Supplementary Information). The observations obtained with the two different instruments are very consistent, with polarization (after correction for interstellar polarization) in HIPPI's reddest band being $p$ = 35.4±1.6 ppm at $\theta$ = 77.1±1.3 degrees, and that measured by PlanetPol being $p$ = 42.1±1.2 ppm at $\theta$ = 79.9±0.7 degrees. The increase in $p$ with wavelength is as expected, and the position angles are very similar. These data were obtained 10 years apart – with PlanetPol in 2005-06, and HIPPI in 2016. We have observed several stars at similar distances and in the same region of the sky with both instruments and none show polarizations anywhere near as high as those seen in Regulus. Furthermore when we fit the position angle of the polarization by minimizing U/I (see Methods) it aligns well with the rotational axis of Regulus as determined by interferometry[3] – shown with green dashed lines in Figure 1.

To provide further confirmation of the mechanism of polarization and to allow constraints on the properties of the star, we have developed a new set of models for the polarization of rotating stars, verifying the code by comparison with the model results of Harrington[18] and Sonneborn[10] (see Methods).

Each model uses as a starting point a value for the angular velocity expressed as $\omega/\omega_c$ where $\omega_c$ is the critical angular velocity for breakup of the star. Given this and a chosen value for the effective temperature ($T_p$) and gravity ($g_p$) at the pole we can calculate the distribution of local effective temperature, $T^l_{eff}$ and $g$ over the star. The effective gravity $g$ is a function of the colatitude $\Theta$ and can be determined by methods described in the literature[2,9]. The variation of $T^l_{eff}$ with colatitude depends on adopting a gravity-darkening law. For our Regulus models we use the gravity-darkening law of Espinosa Lara and Rieutord[19]. This law has been developed specifically for rapidly rotating stars and provides results that agree well with two-dimensional models of rotating stars as well as with interferometric observations, which show generally less variation of $T^l_{eff}$ over the stellar surface than predicted by the classic von Zeipel law[20].

We then create a set of 46 ATLAS9[21] stellar-atmosphere models each calculated for the pair of $T^l_{eff}$ and log $g$ values that apply at colatitudes from 0 to 90 degrees with 2 degree intervals. We calculate the emergent specific intensity (radiance) and polarization as a function of $\mu$ (cosine of the viewing zenith angle) and wavelength for each of these models using a version of the SYNSPEC spectral-synthesis code[22] that we have modified to do radiative transfer with full polarization using the Vector Linearized Discrete Ordinate (VLIDORT) code from RT Solutions[23].

We then select an inclination at which to view the star, and set up a grid of "pixels" covering the observed view, spaced at intervals of 0.01 of the polar radius. For each pixel that overlaps the star we calculate the colatitude $\Theta$ and viewing angle $\mu$. We can then interpolate in our grid of 46 models and 21 $\mu$ values to obtain the specific intensity and polarization for that pixel. These can then be plotted to create images such as those in Figure 2 for a specific wavelength range, or integrated across the star to provide the flux and polarization spectrum at high resolution.

The polarization wavelength dependence is a complex combination of effects arising from the change in polarization as a function of $\mu$ (e.g. Supplementary Figure 4) and the change in relative contributions of equatorial and polar regions at different wavelengths (e.g. Figure 2) and temperatures according primarily to the Planck function[9]. Figure 3 shows how the modeled parameters affect the polarization wavelength dependence. All four main parameters are significant. Increasing the polar temperature ($T_p$) shifts the whole curve upwards, and therefore shifts the crossover from negative to positive polarization to shorter wavelengths. Decreasing the inclination reduces the positive polarization at red wavelengths while having little effect on the negative polarization in the blue. Increasing $\omega/\omega_c$ or decreasing log $g_p$ increases all polarization, making the red polarization more positive, and the blue more negative. As a result, the effects of $\omega/\omega_c$ and gravity cannot be readily distinguished.

It is not, therefore, possible to uniquely determine all four parameters from polarization data alone. To progress further we need additional constraints. Using an inversion of the method for finding distance described by Howarth and Smith[24], we can use the known distance and hence absolute magnitude ($M_V$ = –0.58) and spectroscopically determined projected equatorial rotational velocity, $v$ sin $i$, for which we use 318±8 km/s (see Methods), to derive the global effective temperature, $T_{eff}$ (and hence $T_p$) and log $g_p$ given values of $\omega/\omega_c$ and $i$. This provides a two-dimensional parameter space which we have explored by calculating a grid of models, in the range $\omega/\omega_c$ = 0.9 to 0.99 and $i$ = 65 to 90 degrees; Figure 4 shows the results of fitting these models to our data.

From Figure 4, and after additionally taking account of errors in $v$ sin $i$ and $F_{UV}$ (see Methods), we obtain $\omega/\omega_c$ = $0.965^{+0.006}_{-0.008}$ and $i$ > 76.5 degrees with a best-fit value of 80.0 degrees, which corresponds to $T_p$ = $14375^{+215}_{-181}$, and log $g_p$ = $3.992^{+0.022}_{-0.026}$. The results are in excellent agreement with interferometry[3] where $\omega/\omega_c$ = $0.962^{+0.013}_{-0.028}$ and $i$ = $86.3^{+1.7}_{-2.3}$ degrees was obtained. The bottom left panel of Figure 3 shows that polarization is particularly sensitive to $\omega/\omega_c$ near critical velocity, consequently we are able to constrain $\omega/\omega_c$ more tightly than interferometry. Our determination of inclination is not as precise as that from interferometry. It can be seen from Figure 3 that the models are not very sensitive to inclination near 90 degrees. This difficulty would not be encountered if the star were less inclined; in the present instance it might be overcome by high-precision spectropolarimetry. Whilst $U/I$ is negligible across the broad band, our models show significant polarizations in spectral lines due to the Öhman Effect[25] (see Supplementary Information). This polarization is zero at 90 degrees and increases rapidly with decreasing inclination.

Particularly in extreme stars such fundamental parameters as we've determined here are of principal importance. They are needed to, for instance, accurately track the star's evolutionary path along the H-R diagram[5] and therefore its end state (e.g. supernova; compact remnant), or to determine the limits of the Be phenomenon. Rotation also impacts how much mass is lost back into the ISM and the local space weather in the form of stellar winds[4]. Yet, to date only half a dozen rapidly rotating stars have been resolved with interferometry, whereas our initial estimates indicate we may be able to determine parameters for thrice that number with our present instrumentation and facilities. This result thus represents a watershed moment for stellar linear polarimetry. Previously the field has been largely restricted to studying material external to stars or those with extreme magnetic fields[12]. Now we are able to probe fundamental parameters of the stellar atmosphere itself.

**Acknowledgments:** The work was supported by the Australian Research Council through Discovery Project grants DP140100121 and DP160103231.We wish to acknowledge Daniela Opitz, Jacob Sturges and the staff at the AAT for their assistance in making the HIPPI observations. We thank Robert Spurr of RT Solutions for providing the VLIDORT software. We would also like to thank the anonymous referees for their thoughtful feedback.

**Author contributions:** J.B., D.V.C., L.K-C., K.B., P.W.L. and J.H.H. drafted the initial proposals to observe Regulus with HIPPI, this followed an extended discussion of the PlanetPol results dating back to 2010 that included I.D.H., J.H.H., J.B. and P.W.L. All authors contributed to the discussion and drafting of the final manuscript. The HIPPI observations were carried out and directed by J.B., D.V.C., L.K-C. and K.B. In addition the following authors made specific contributions to the work: D.V.C. contributed the initial data analysis, telescope polarization subtraction, the stellar atmosphere modeling, the interstellar subtraction, model comparison to data and other calculations including the position angle calculation and the calculations related to Regulus' companions. J.B. contributed the polarized radiative transfer modeling and verification, gravity darkening calculations and PlanetPol bandpass model calculations and code. I.D.H. contributed rotational velocity calculations, the knowledge and calculations needed to constrain parameter space, and other miscellaneous calculations. K.B. contributed HIPPI bandpass model calculations and code, and research on Regulus' companions. P.W.L. contributed details of the PlanetPol observations not otherwise available.

**Competing interests:** The authors declare no competing financial interests.

...

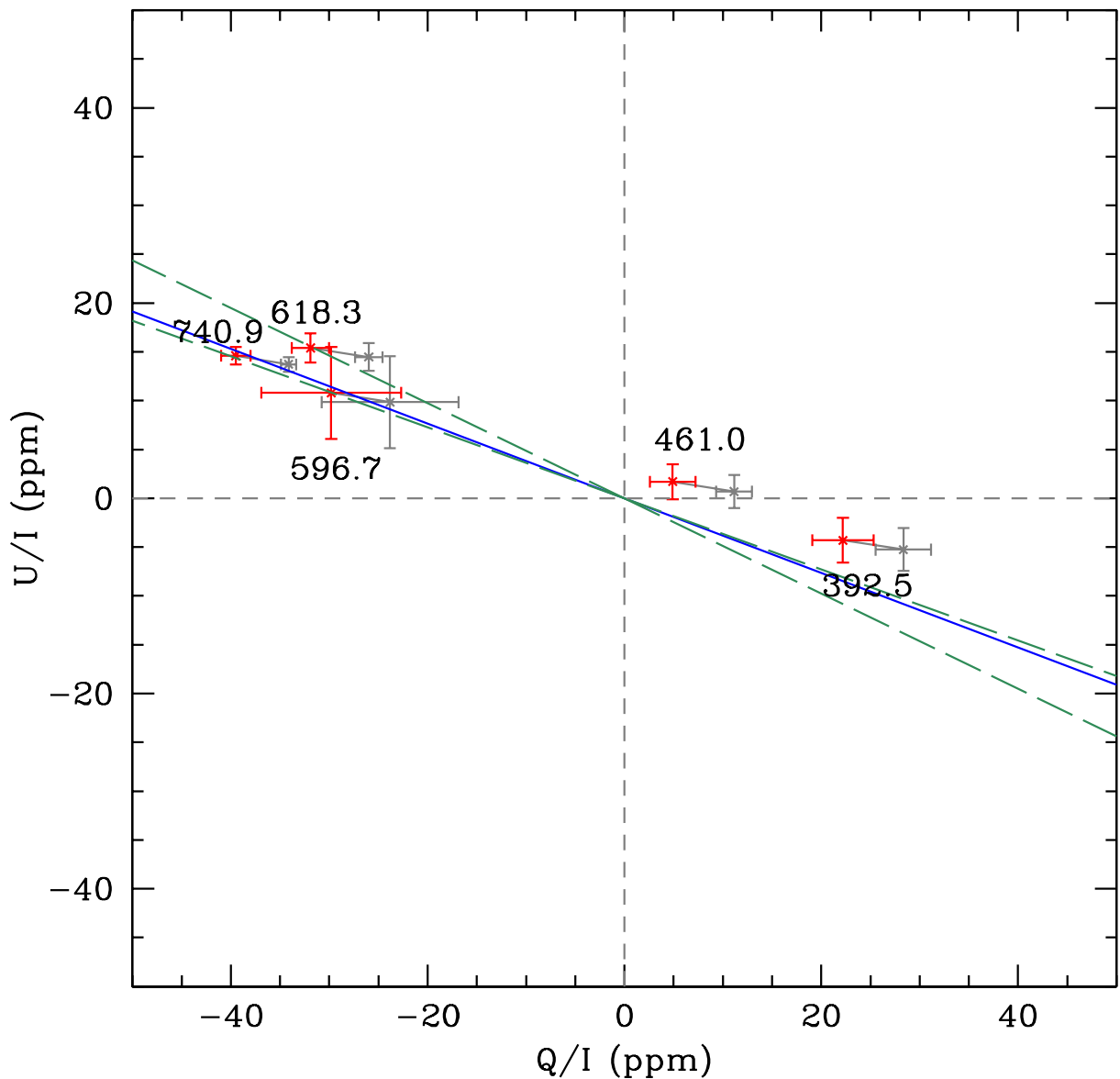

**Figure 1 | Polarization observations of Regulus on a Q-U diagram.** Polarization observations of Regulus averaged for each band (grey points) and the same observations corrected for interstellar polarization (red points) – 1-sigma error bars for the observations are derived from the internal measurement statistics and those of the telescope polarization; the interstellar corrected points also include a contribution derived from the scatter in interstellar polarization measured in nearby stars. Labels show wavelength in nm. The blue line is the fitted orientation for the rotation axis of the star (see Methods). The green dashed lines show the bounds on the rotation axis of the star as determined by interferometry[3].

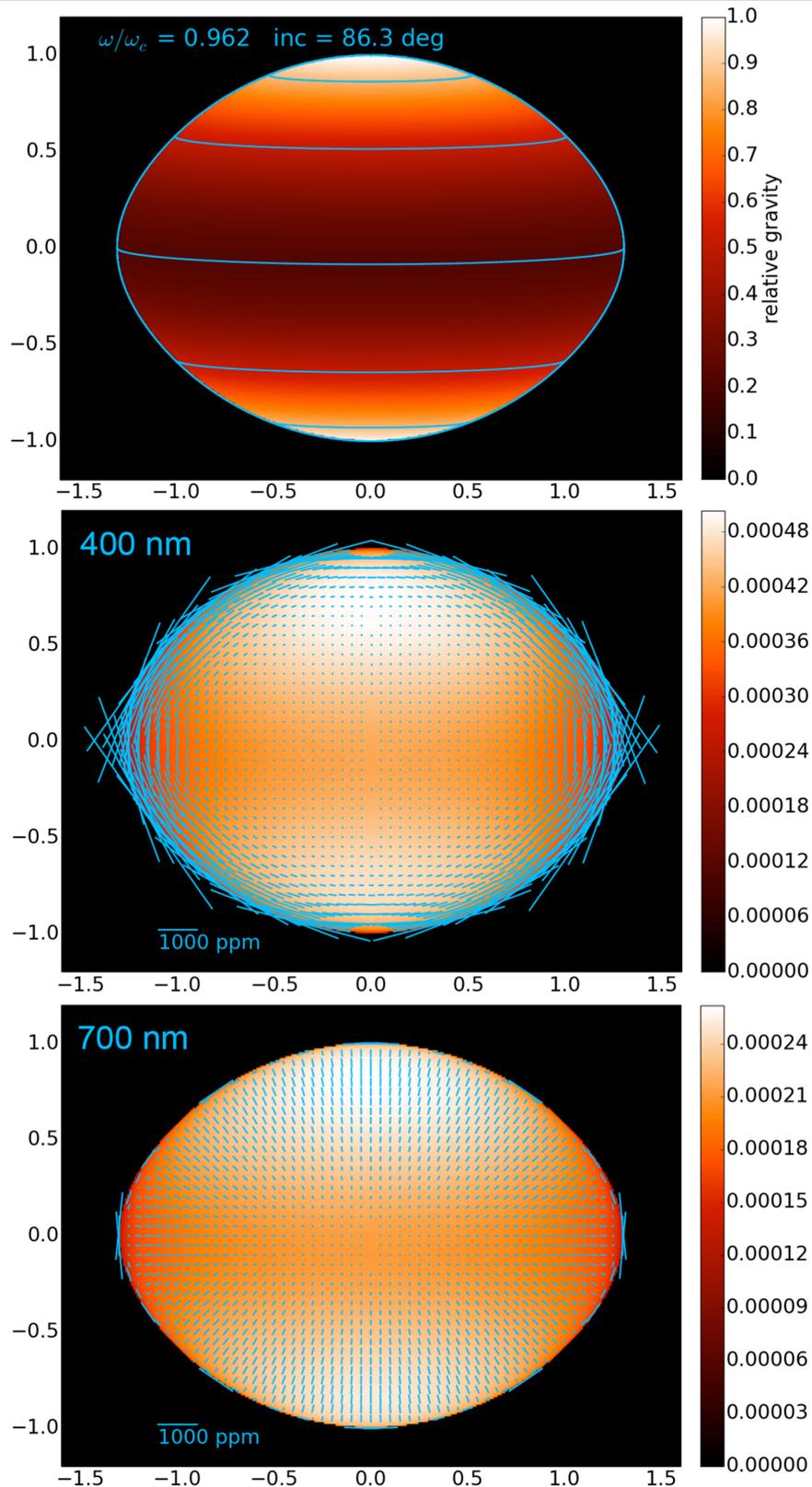

**Figure 2 | Example of polarization modeling of Regulus.** The upper panel shows the distribution of effective gravity over the surface of the star, in this case for $\omega/\omega_c$ = 0.962 and inclination *i* = 86.3 degrees (the interferometry parameters determined by Che et al.[3]). The lower panels show the derived intensity distribution over-plotted with polarization vectors at wavelengths of 400 nm and 700 nm. The intensity scale is specific intensity (or radiance) $I_v$ in units of $erg.cm^{-2}.s^{-1}.Hz^{-1}.sr^{-1}$.

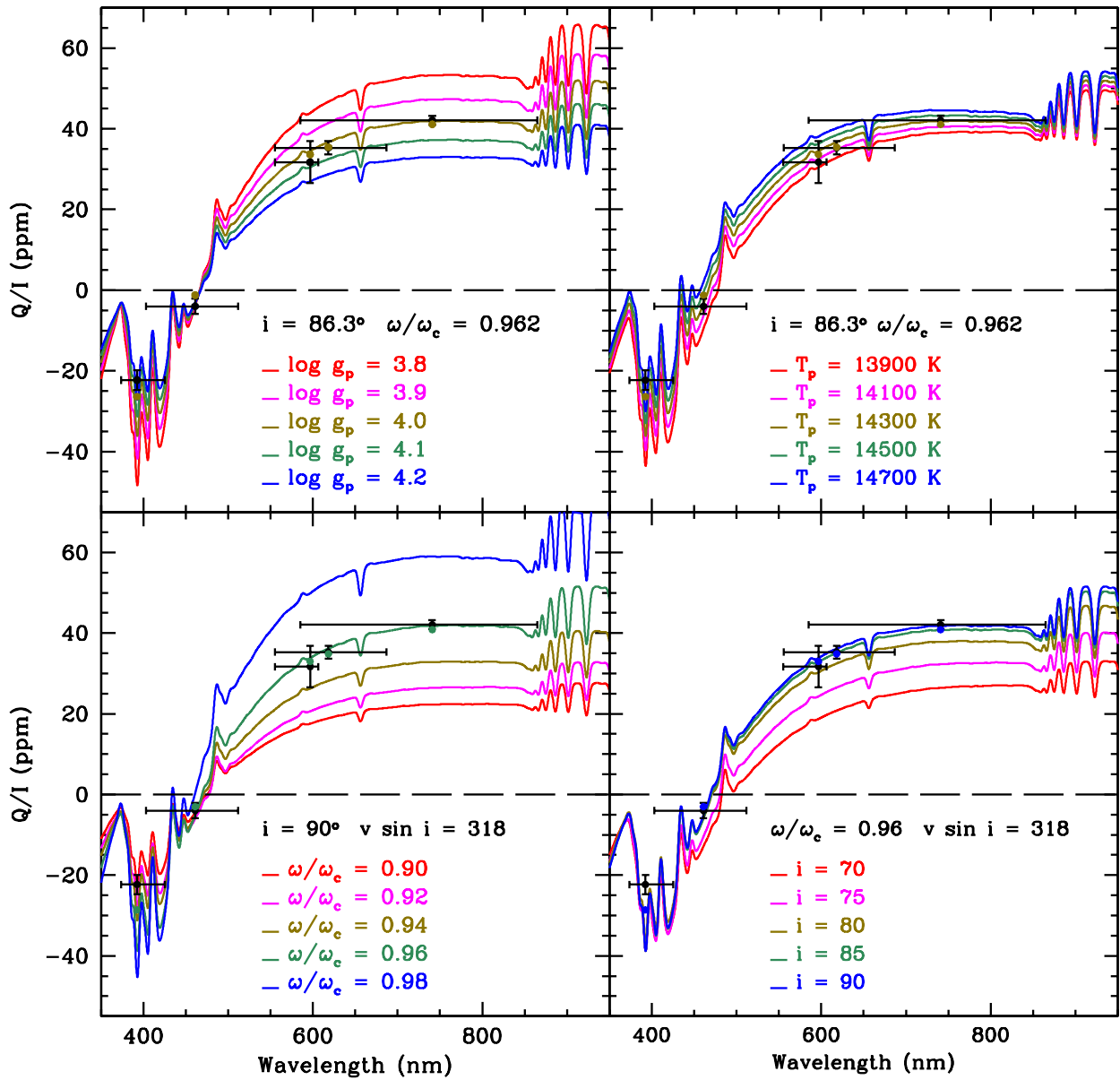

**Figure 3 | Observations of Regulus vs. model results.** Observations of Regulus (black points with error bars) compared with smoothed model results showing the effect of varying different parameters (polar gravity, polar temperature, rotation and inclination). The observations have been rotated into the plane of Regulus' determined rotational axis (i.e. by 79.5 degrees), and the error bars show both the formal error in *Q/I* (as per Figure 1) and the full-width-half-maximum of the spectral bands. Solid colored circles show the best-fit model averaged over the observed bandpass. The upper two model sets use the interferometrically determined parameters $\omega/\omega_c$ = 0.962 and inclination *i* = 86.3. The lower two models use parameters for a fixed *v* sin *i* of 318 km s$^{-1}$ as explained in the text.

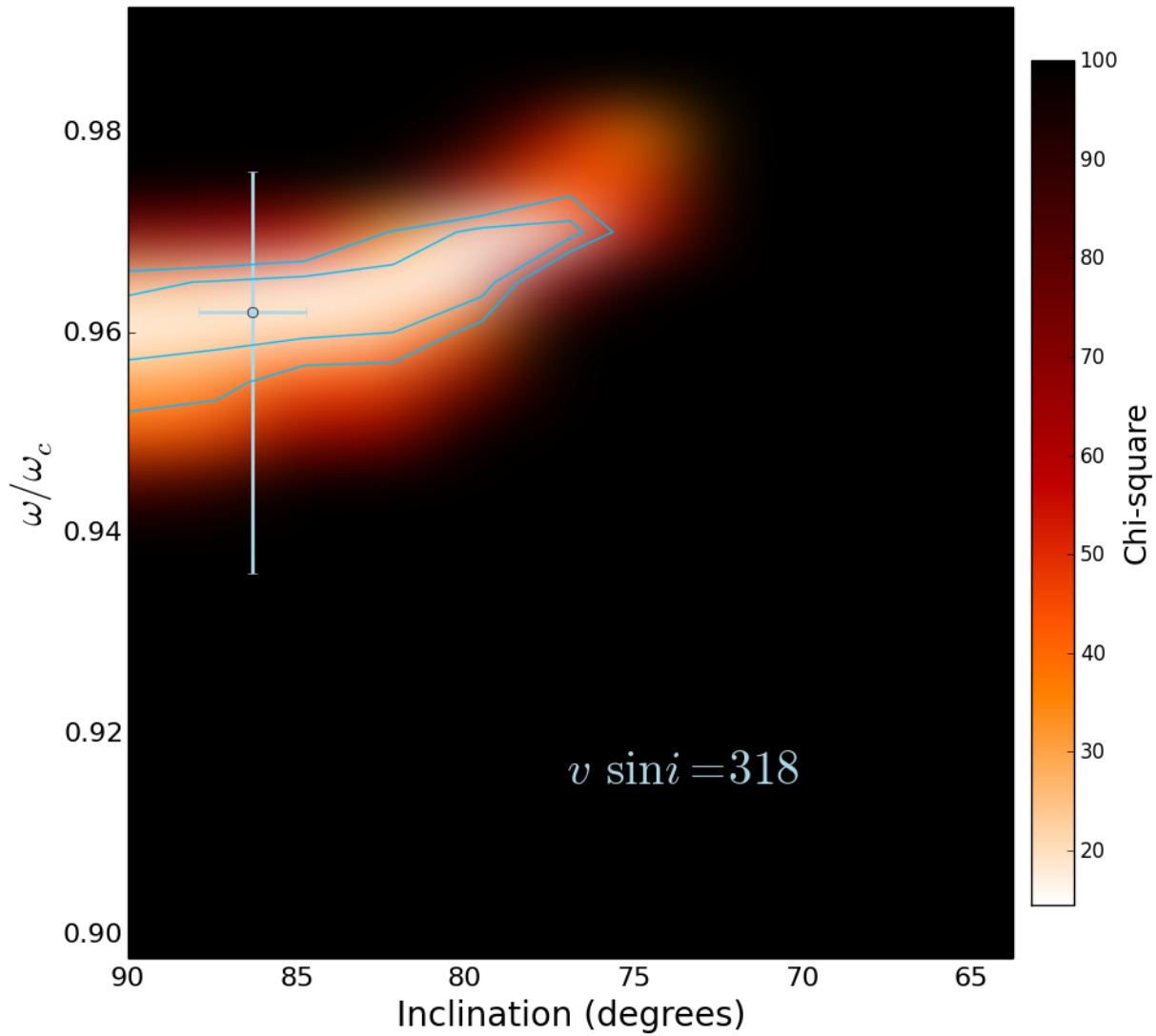

**Figure 4 | Results of model fitting.** Results of a grid of models calculated for *v* sin *i* = 318 km s$^{-1}$ fitted to the polarization observations as a function of wavelength. The resulting $\chi^2$ is plotted as a function of rotation rate and inclination. Models near ω/ω$_c$ = 0.96 at 90 degrees, 0.965 at 80 degrees, or slightly higher values for lower inclinations are the best fits. The contour lines mark the formal 1-sigma and 2-sigma uncertainties using a reduced $\chi^2$ statistic. The point with error bars shows the parameters from interferometry and the associated 1-sigma uncertainty[3].

## Methods

**Multi-band high-precision polarimetry.**

We made observations with two high-precision polarimeters – PlanetPol[15] and HIPPI[14]. High-precision polarimetry requires the use of high-frequency modulation to eliminate the effects of astronomical seeing. PlanetPol uses Photo-Elastic Modulators (PEMs) oscillating at tens of thousands of Hertz. HIPPI uses Ferro-electric Liquid Crystal (FLC) technology to provide modulation at 500 Hz.

The PlanetPol (PP) observations were made in a single broad red band that runs from 575 to 1025 nm. With HIPPI we utilised three different filters, those being a 425 nm short-pass filter (425SP), and SDSS g' and r' filters. All three filters were used in combination with the blue-sensitive Hamamatsu H10720-210 ultra bialkali photocathode PMT (b). We also made observations with the r' filter paired with the red-sensitive Hamamatsu H10720-20 infrared-extended multialkali photocathode PMT (r), resulting in 5 bands in total.

Because of the filter and PMT profiles, and the modulator performance, the effective wavelength and polarization efficiency varies with the spectral type of the target. We have calculated these for each band using the bandpass model described by Bailey et al.[14]. Some, but not all, of these have been reported previously[14,26,27], so for completeness we list the bandpass calculations for each filter/PMT combination used here in Supplementary Table 1, along with the corresponding PlanetPol bandpass calculations given by Hough et al.[15]. The stars we observed for this work – Regulus and various calibration standards – are within 60 pc, and so no interstellar reddening has been applied in the bandpass model. We have also made bandpass calculations specifically for Regulus. The stellar model used for this is amongst the suite of models described in the Polarization Radiative Transfer section; specifically it is the one with $\omega/\omega_c = 0.95$ and $i = 90$ deg. The various band sensitivities and effective wavelengths as applied to the Regulus stellar model are shown in Supplementary Figure 1.

The r' (b) observations in May 2014 used a different modulator to all the other HIPPI observations. This gave a 1.5% higher efficiency than the data in Supplementary Table 1 (see notes in the table).

**Observations and calibration.**

The PlanetPol observations, taken in April 2005 and February 2006 at the 4.2 m William Herschel Telescope (WHT) in La Palma, have been detailed previously[17]. There the standard A0 efficiency correction from Supplementary Table 1 was applied[17]. Here we have instead applied a bandpass model specific to Regulus, and error weighted the two observations; the details are otherwise unchanged.

The HIPPI observations were acquired in four observing runs over three years, on the Anglo-Australian Telescope (AAT) at Siding Spring in Australia and are reported here for the first time. For each observation we took measurements at four different position angles of the AAT's Cassegrain rotator (0, 45, 90 and 135 degrees). The effects of the background sky are typically removed through the subtraction of a sky measurement, obtained at a separation of 2 arcmin for each PA setting and for each observation. The duration of the sky measurements was 3 minutes per Stokes parameter. The observing, calibration and data-reduction methods are described in full detail by Bailey et al.[14].

Observations of high-polarization standards ($p$~1-5%) were used to calibrate position angle. A list of the high-polarization standards observed (in the g' band but occasionally checked against similar measurements in other bands) is given in Supplementary Table 2. The precision of each determination is taken to be ±0.5 degrees, based on the mean consistency of the calibration provided by the different reference stars, which themselves have uncertainties of this order.

We use observations of stars previously measured with negligible polarizations to determine the zero-point or telescope polarization (TP). Deviations from zero polarization in the standards chosen can therefore be a source of error. To minimize this complication we keep a very short list of low-polarization standards, about which we can be most certain. Even so, variance in the low-polarization standards could produce differences up to a few ppm in the assumed telescope zero point between runs. To combat this problem here – where exquisite precision is required – we have assigned each low-polarization standard a non-zero polarization based on the best available data.

The assigned polarization values for each standard star are given in Supplementary Table 3, for each band in which they were observed. The assigned values and associated errors for α Ser, β Leo, and β Vir come from PlanetPol observations[17], and are adjusted for other bands based on the polarimetric color of the ISM (explained in more detail in the next section). By assuming only interstellar polarization we can employ the method outlined in Cotton et al.[28] to determine values for Sirius and β Hyi; this method is described in full in its application to the interstellar subtraction for Regulus in the next section. QU plots made to inter-compare our low-polarization measurements are sparsely populated but seem to corroborate the assignments.

The assigned values are subtracted from each standard measurement, and their uncertainties propagated. The error-weighted mean of the standard measurements is then used determine the TP. For the error in TP we use the greater of either the root-mean-square propagated errors, or the error-weighted mean of the scatter (standard deviation) in the individual measurements, whichever is greater. This methodology has its greatest consequence for the May 2014 observations, as there is a significant difference between measurements of α Ser and β Leo. β Leo hosts a debris disk[29], and while the polarization determined with PlanetPol is suitably small, there may be an intrinsic component resulting in anomalous deviation at shorter wavelengths. Subsequent observations of α Ser suggest very-low-level variability may be present; our data are as yet insufficient to test our suspicions. The May 2014 run is the only one where α Ser and β Leo were used as standards in this work. In other subsequent work we have tried to avoid using either of these stars as standards. Given these circumstances, the larger error assigned according to the scatter is appropriate.

The observations of the low-polarization standards, rotated into the QU-frame and corrected according to their assigned values (from Supplementary Table 3) and the resulting TP are given in Supplementary Table 4.

Observations of Regulus were made during three different observing runs. The exposure times varied according to target availability, weather and other considerations. The observations lasted approximately half an hour to an hour, with second-stage chopping employed to swap the sign of the modulation every 40 to 80 seconds. The observations are detailed in Supplementary Table 5. We also observed HD 99028 as an interstellar control; the details of this along with the PlanetPol observations are also given in Supplementary Table 5.

**Interstellar polarization subtraction.**

Within the Local Hot Bubble (LHB) interstellar polarization is small (roughly 0.2 – 2 ppm/pc). Yet even at this level accurate subtraction is crucial to measuring intrinsic polarization induced by rapid rotation. Only recently has sufficient precision been obtained in measurements of stars within the LHB[17,26,28,30] to enable accurate interstellar subtraction in this region.

To determine the contribution to our measurements from interstellar polarization we have, in the first instance, used the same method as in past work[28]. Regulus is north of Galactic latitude b = +30. From measurements of around 30 stars, the relationship between the magnitude of polarization and distance for this part of the ISM, in the g' band, has been determined, on average, to be[28] $p_i = (0.261\pm0.017)d$, where $p_i$ is in ppm and $d$ is in pc. Regulus lies at a distance of 24.3 pc[31], which gives $p_i = 6.3\pm0.4$ ppm. Because the interstellar medium can be clumpy we have adopted a more conservative estimate of the error in $p_i$, which we describe below.

Interstellar polarization varies as a function of wavelength. The best current estimate of interstellar polarization color within 100 pc is described by Marshall et al.[27] as likely peaking at 470 nm, having been calculated using the empirically-determined Serkowski Law[32] as modified by Wilking[33]; we have used the relation given there to adjust the magnitude of interstellar polarization in the other bands using their effective wavelengths. The most extreme difference is between the g' and PlanetPol bands and is ~1 ppm.

In nearby space it has been shown[28] that the position angles of pairs of stars polarized by the interstellar medium are correlated at angular separations up to 35 degrees; with the correlation being greater at closer separations. In that work an angular separation weighted average of intrinsically unpolarized A-K stars within 50 pc of the Sun was used to determine interstellar polarizations for nearby FGK dwarfs. Here we use the same approach and list of stars (with the addition of HD 99028) but extend the distance limit to 70 pc from the Sun, since Regulus lies at a greater distance than most of the targets in the earlier study. The control stars from the list with angular separation to Regulus less than 35 degrees are listed in Supplementary Table 6. For each of these stars we have determined a debiased polarization with distance (where the debiasing is a standard way to account for the effects of measurement error on the positive definite quantity *p*), and calculated the equivalent polarization at Regulus' distance. The error-weighted scatter in these measurements from the determined interstellar polarization (i.e. 6.3 ppm) has been taken as the error in our interstellar determination for Regulus; the adopted value thus being 6.3±1.4 ppm.

The position angles of all the assumed intrinsically unpolarized stars near Regulus are shown in Supplementary Figure 2. Within 35 degrees three stars meeting the required criteria were observed with PlanetPol[17], another with HIPPI in the g' band during a bright-star survey[26] and we have observed one further star with HIPPI in the g' band; these are listed in Supplementary Table 6. Taking the separation-weighted average of the position angles of these stars gives the interstellar position angle at Regulus as 175±12 degrees, where the weight was calculated as 1-(*s*/35), where *s* is the angular separation to Regulus in degrees. We assume no angular dispersion with wavelength and subtract the interstellar polarization at the same angle in every band. Our Regulus observations before and after interstellar subtraction are shown in Supplementary Table 7.

**Interstellar reddening.**

In our bandpass model and in the stellar modeling that follows we adopt $E_{(B-V)} = 0$. This assumption is justified on the basis of the very low level of interstellar polarization. Within the LHB, photometry is not sensitive enough to accurately measure interstellar reddening[17]. However, past studies have looked at the correlation between $E_{(B-V)}$ and $p_i$ beyond the LHB, and we can use these to infer a likely range for $E_{(B-V)}$ here. A summary of these studies given by Clarke[12], who mentions extremes of $p_i/E_{(B-V)} \leq 9.0$[32] and $0.18$[34]. Although we do not know what the specific relationship is for nearby space, it is reasonable to conclude that $E_{(B-V)}$ is probably much less than 0.005.

**Determination of Regulus' rotation-axis position angle.**

A determination of Regulus' rotation-axis position angle, $\theta_R$, can be made after subtraction of the interstellar polarization from the raw measurements. This is accomplished by assuming any intrinsic polarization is due to rotational oblateness, and rotating $Q/I$ and $U/I$ into a new reference frame ($Q_r/I$, $U_r/I$) that minimizes $U_r/I$. In practice we use Python's 'scipy.optimize.curve_fit' routine with its 'trf method' and error weight each of the bands. This gives $\theta_R$ as 79.5±0.7 degrees in good agreement with the interferometric determination of 77 to 80 degrees[3]. The polarization in each band after rotation by this angle is given in Supplementary Table 7.

**Determination of projected equatorial rotation velocity.**

As discussed below, Regulus' equatorial rotation velocity, *v* sin *i*, provides a useful constraint on system parameters. The literature reports a rather wide range of values for the former parameter (~260 – 350 km/s; e.g.,[35,36]), so to make a new determination we de-archived high-resolution spectra obtained with several echelle spectrographs (UVES[37], HARPS[38], FEROS[39], and Elodie[40];

resolving powers ~5 – 10x10$^4$, signal:noise ratios ~500 for each instrument in merged datasets). The spectrum of Regulus is dominated by strong hydrogen-Balmer lines (consistent with its spectral type), with the star's rapid rotation leading to blending of other characteristic features (specifically, the HeI/MgII 447.1/448.1 nm classification lines). After examining the spectrum, we chose the isolated, unsaturated Si II 634.7 nm line for analysis (as did Gavrilović & Jankov[41]). This is a 'cold' line (its equivalent width increases monotonically going from equator to pole in our models), and continuum gravity darkening is not severe at this wavelength (half the observed 650 nm continuum flux originates from colatitudes ≳ 70 degrees); consequently Si II λ634.7 does not suffer strongly from the degeneracy between line width and equatorial rotation velocity discussed by Townsend, Owocki & Howarth[42].

While direct modeling of the spectral-line profile can yield estimates of $v \sin i$, its Fourier transform encodes this information in a more straightforwardly accessible manner, with the first minimum in the FT scaling linearly with $v \sin i$ for a spherical star and wavelength-independent limb darkening (e.g.[43]). However, rotational distortion, gravity darkening, and wavelength-dependent limb darkening all modify the scaling factor, and noise can influence the precise location of the minimum. Synthetic spectra were therefore generated for full rotating-star models, using Roche geometry and ATLAS9 model atmospheres as described below, for an appropriate range of $v \sin i$ values. (The results are insensitive to the precise values adopted for other stellar parameters, as we confirmed through sensitivity tests.) The model spectra were convolved with approximate (Gaussian) instrumental profiles; and noise, determined directly from the observations, added. The Fourier transforms of the observed and model spectra were then compared in order to determine $v \sin i$. The modeling fully accommodates any small residual nonlinearity between line width and rotation velocity, and we verified, through sensitivity tests, that the results are insensitive to the precise values adopted for other parameters (inclination, effective temperature, polar gravity, and $\omega/\omega_c$).

The Fourier transforms of the observed and model spectra were then compared in order to determine $v \sin i$. We estimate $v \sin i \approx$ 336, 318, 310, 318 km/s from the Elodie, FEROS, HARPS, and UVES spectra, respectively, and adopt 318±8 km/s (where the estimated error is dominated by the inter-instrument differences, and accommodates both the small individual uncertainties and the weak sensitivity to other parameters). Comparison of observed and model spectra confirms that this value leads to a satisfactory match, and our result is consistent with the other direct measurements published this century of 300±20 km/s[41], and 317±3 km/s[5].

**Constraining stellar parameter space.**

Any modeling of rotating stars (including our polarization calculations) requires a number of stellar parameters as inputs: three to characterize the basic structure (e.g, effective temperature, mass, and polar radius, or surrogates) and two that describe the rotation (e.g., $v \sin i$ and inclination, $i$, or surrogates).

The effective temperature of Regulus is such that the bulk of its radiant-energy output is directly observable, and hence the spectral-energy distribution gives a strong constraint on the global effective temperature $T_{eff}$. Furthermore, the distance to Regulus is precisely determined from *Hipparcos* astrometry[31]: $d$ = 24.3±0.2 pc. Given $T_{eff}$, the apparent brightness gives the angular size of the star as seen from Earth, whence $d$ yields the stellar radius (e.g., polar or equatorial radii $r_p$, $r_{eq}$), for any given values of $\omega/\omega_c$ and $i$.

Given $\omega/\omega_c$ and $i$, the observed $v \sin i$ establishes a value for $\omega_c$, which in turn yields the stellar mass (within the framework of the Roche model; see, e.g., Howarth & Morello[44]) for the corresponding $r_{eq}$. Although the line profiles and flux distribution (especially the size of the Balmer jump) potentially offer further leverage on parameters such as polar gravity, $g_p$, in practice observational uncertainties limit the additional utility of those diagnostics.

These principles can therefore be used to determine the effective temperature and polar gravity (along with the stellar mass and radius) from the known distance plus the observed flux distribution and projected equatorial rotation velocity, given values for $\omega/\omega_c$ and $i$ (which are constrained by the polarization modeling described below).

In practice, we determined temperatures and radii, for specified values of $v \sin i$, $\omega/\omega_c$, and $i$, by requiring rotating-star model fluxes to reproduce simultaneously the observed magnitude in the Johnson *V* band (effective wavelength ~550 nm) and the UV flux integrated over the range 132.5 – 194.5 nm. The UV flux was measured from the archival low-dispersion spectrum SWP 33624 (the same spectrum used by McAlister et al.[5]), obtained with the *International Ultraviolet Explorer* satellite and downloaded from the IUE Final Archive at the Mikulski Archive for Space Telescopes. The required model-atmosphere fluxes and magnitude normalization were taken from Howarth[45] and are consistent with those used in the polarization calculations.

The projected equatorial rotation velocity is quite well determined, leaving $\omega/\omega_c$ and $i$ as the principal 'free' parameters. Supplementary Figure 3 shows results of an initial grid, sampled at 0.01 in $\omega/\omega_c$ and every 5 degrees in $i$. We calculated additional points on a finer grid (0.005 in $\omega/\omega_c$ and 2.5 degrees in $i$) around the region of the best-fitting solution from the polarization analysis. Final numerical results of our analysis of the basic stellar parameters are incorporated into Supplementary Table 8 (where the uncertainties are generated from Monte-Carlo simulations, sampling the estimated errors in the input parameters).

**Modeling.**

All our modeling is based on the assumption of a Roche model and uniform rotation. On this basis the geometry of a rapidly-rotating star follows equations given by Harrington & Collins[2] that specify how the radius and other properties vary as a function of colatitude ($\Theta$) and rotational velocity expressed as $\omega/\omega_c$. Each colatitude has a different value of gravity ($g$) and local effective temperature ($T'_{eff}$). To represent this variation we create for each model (where a model is defined by its polar temperature $T_p$ and gravity, its $\omega/\omega_c$ and its gravity darkening law) a set of 46 ATLAS9 model atmospheres representing the $T'_{eff}$ and $\log g$ of the star at 2 degree intervals of colatitude from 0 to 90 degrees. The ATLAS9 models use as a basis the Castelli & Kurucz model grid[46] for solar abundances[47].

**Polarized radiative transfer.**

To calculate the polarization of the emission from these models we use a modified version of the SYNSPEC spectral synthesis code of Hubeny – briefly described by Hubeny, Stefl and Harmanec[22]. We have modified the code to do polarized radiative transfer using the Vector Linearized Discrete Ordinate (VLIDORT) code[23] which is a comprehensive implementation of the discrete-ordinate method of radiative transfer.

From the emission, absorption and scattering properties of each layer of the atmosphere available in SYNSPEC it is straightforward to provide the inputs needed by VLIDORT for each atmospheric layer. We assume that all scattering is described by a Rayleigh scattering matrix, which is appropriate for Thompson scattering from electrons. Our modified version of SYNSPEC outputs the intensity and polarization as a function of viewing angle ($\mu$ = cosine of the local zenith angle) and wavelength. SYNSPEC uses a non-uniform wavelength scale in order to fully sample line structure, which we re-bin to a uniform 0.01 nm spacing for subsequent processing.

To apply the model to a rotating star we use the geometric model to determine the colatitude $\Theta$ and $\mu$ as a function of coordinates $x$, $y$ in the observed image plane for a given inclination $i$. We divide the observed image into a large number of "pixels" using a spacing of 0.01 in units of the polar radius of the star. We then interpolate in the grid of models to obtain intensity and polarization for each pixel. We use linear interpolation in $\Theta$ and cubic spline interpolation in $\mu$. The polarizations require rotation through an angle $\xi$ as defined by Harrington and Collins[2] to rotate into a coordinate system defined by the rotation axis of the star.

The results of this process are intensity and polarization values for each pixel, at the full spectral resolution of the model, that can be plotted as an image and vector map as in Figure 2, or can be integrated over the star to give the total polarization. Because we work in the observed image plane and use a large number of pixels (typically about 40,000) the integration is a simple sum over pixels for each Stokes parameter. To affect a comparison to observation, as in Figure 4, the integrated intensity and polarization is fed into the bandpass model at the full spectral resolution of the model.

**Tests of the modeling.**

We have tested our polarization version of SYNSPEC by calculating polarization for a number of stellar-atmosphere models that are included in the results of Harrington[18]. Supplementary Figure 4 shows an example for TLUSTY[48] models for $T_{eff}$ = 15000 K and log $g$ = 2.5 and 3.5 at a wavelength of 400 nm. The polarizations from the two models agree very well. There is a small difference in specific intensity between Harrington's results and those from SYNSPEC/VLIDORT but our results agree very well with intensity from SYNSPEC's standard radiative transfer. This is probably due to the fact that Harrington's calculations are for continuum only, whereas our models include spectral lines. Similar agreement has been found for other wavelengths and different temperature models.

To test our rotating-star models we have compared with results reported by Sonneborn[10]. Supplementary Figure 5 shows a comparison of our modeling with Sonneborn's results for B1, B2 and B5 stars with $\omega/\omega_c$ = 0.95 and $i$ = 90 degrees. Sonneborn does not specify the temperature and gravity values used for these models. However, we found that we could match his results closely with $T_p$ = 28000 K (B1), 23000K (B2) and 18000 K (B5) with a log $g_p$ of 4.07 in all cases. For these models we used a von Zeipel gravity darkening law[20] consistent with Sonneborn's original work. In a future work we plan to examine stars with different spectral types in more detail.

**Regulus final parameters.**

The result of fitting our model grid to the data is shown in Figure 4. From this we obtain a best fit in $\chi^2$ and the 1-sigma confidence region in reduced $\chi^2$ that constitutes the model error. To obtain the final errors we must also consider the errors in $T_{eff}$ and log $g_p$ due to the uncertainty in the parameters obtained from the literature, significantly, the UV flux, $F_{UV}$, (which we take as 5%) and $v \sin i$. These are shown diagrammatically in the bottom left corner of Supplementary Figure 3 and are simply added as root-mean-squares to the model errors to obtain the final uncertainties in $T_{eff}$ and log $g_p$. To calculate the errors in $\omega/\omega_c$ and $i$ the observational uncertainties in $T_{eff}$ and log $g_p$ must first be transformed into a corresponding change in polarization via an interpolation of the upper two panels of Figure 4; the change in polarization can then be equated to errors in $\omega/\omega_c$ and $i$ through the lower two panels of the same figure, and finally added as root-mean squares to the model errors. Supplementary Table 8 lists all the parameters adopted and calculated in modeling Regulus, and the final values determined. For other derived physical parameters, given at the bottom of Supplementary Table 8, the errors are propagated from the uncertainties on the model-determined parameters using Monte-Carlo methods.

**Data availability:** All processed data generated during this study is included in this published article (and its supplementary information files), the raw data files are available upon reasonable request. All other data analysed in this work comes from public repositories, where this is the case the origin of the data is indicated in the text.

## Supplementary Information

**Potential influence from Regulus' companions.**

According to Gies et al.[49] Regulus has a companion star orbiting with a period of 40.11 days with a probable semi-major axis of 0.35 au – placing it at an angular separation of 0.014 arcsec, easily within the HIPPI and PlanetPol apertures, which are 6.7 and 5 arcsec respectively. The companion has a low mass (> 0.30 $M_{Sun}$) and is thought to be a white dwarf, but might also be a main-sequence M4 star.

If the companion is an M-dwarf the magnitude difference to the primary is thought to be equal to ~6 in the K-band[49]. A simple calculation assuming blackbody emission shows that, in this case, the companion would have to be polarized at more than 3000 ppm to contribute even 1 ppm to the overall signal in the most affected band (that being the PlanetPol band). No broadband linear-polarization measurements of M dwarfs have been made with modern equipment. A review of older measurements of red dwarfs given by Clarke[10] relates null detections even for very active stars like BY Dra. Recently we[28] have found active K dwarfs can be polarized, but only, so far, at low levels – up to around 50 ppm. Thus if the companion is an M dwarf we can quite confidently rule out any significant contribution from it.

If the companion is a white dwarf the magnitude difference in the K-band is given as ~10 by Gies et al.[49]. In this case the same calculation as above shows that the 425SP band is the one most likely to be effected; and an intrinsic contribution from the companion of ~2500 ppm would be required to change our measurement by 1 ppm. A recent R-band survey of non-magnetic white dwarfs[50] failed to make any 3-sigma detections, with a detection limit of typically ~5000 ppm. This gives us confidence that a significant intrinsic contribution from the companion is unlikely.

In addition to an intrinsic contribution from the companion itself, material entrained between the two stars could provide a scattering medium that would produce a non-zero signal. Any such signal would be variable on the timescale of the orbital period. We do not have data to specifically test this time-scale, however we regard it as unlikely, in large part because our repeat observations in multiple bands (as given in Supplementary Figure 5) demonstrate a high degree of consistency when considering the precision of the individual measurements.

The double star HD 87884 is also considered to be companion to Regulus, but at an angular separation of 175.94 arcsec[51] it is too distant to influence our observations.

**The Öhman effect and inclination.**

As a result of including the rotational Doppler shift in the polarized radiative transfer modeling, structure can be seen across spectral lines. Öhman[24] first inferred this result by analogy with Chandrasekhar's original modeling paper[1]. Much later, more quantitative results were obtained by Collins & Cranmer[52]. They evaluated the effect for pure absorption spectral lines with a Struve-Unsöld model[53,54] that included limb darkening. For a B type star the magnitude of the effect was found to be 10s of ppm in *Q/I* for lines in the visible region of the spectrum, but much more than that in the UV[52]. Thus far the phenomenon has not been observed. If a high-precision spectro-polarimeter were to be turned to the task greater precision in determining inclination might be obtained. This is because the polarization in the spectral lines varies more strongly with inclination at high inclinations than does the broadband polarization. This can be seen in Supplementary Figure 6, where we show the region from 380 to 450 nm – where the greatest density of strong spectral lines is to be found – in both *U/I* and *Q/I* (where our results are of a similar magnitude to those of Collins & Cranmer in the visible region[52]). Particularly noteworthy is the rapid (positive and negative) increase in *U/I* away from 90 degrees inclination – up to 10 ppm in 10 degrees.

**A comment on differential rotation.**

As described earlier, all the modeling carried out here assumes rigid stellar rotation; this assumption is supported by conclusions in the literature that there is no significant surface differential rotation in early-type stars[23]. Yet, it is worth noting the possibility rapidly rotating early-type stars may not be rigid rotators[55,56]. Examining the effect of differential rotation on polarization is beyond the scope of this work. However, we have run models with and without Doppler velocities included to examine the Öhman effect, with no significant effect on the broadband polarization. Differential rotation will manifest in broadband polarization through changes in the latitudinal temperature distribution and the shape of the star. Both effects we expect to be modest, but we note that the approach we adopt here is well suited to investigating this question, since ultimately we are taking the sum of elements distributed over the disk of the star, and the stellar models representing each of these elements may be adjusted independently to correspond to a differential rotation model.

On the other hand, while line profiles aren't particularly sensitive to the shape of the star, they do map the distribution of temperature more directly than would the broadband polarization. As a consequence stellar spectral lines are a demonstrably better diagnostic of differential rotation[23].

**Supplementary References**

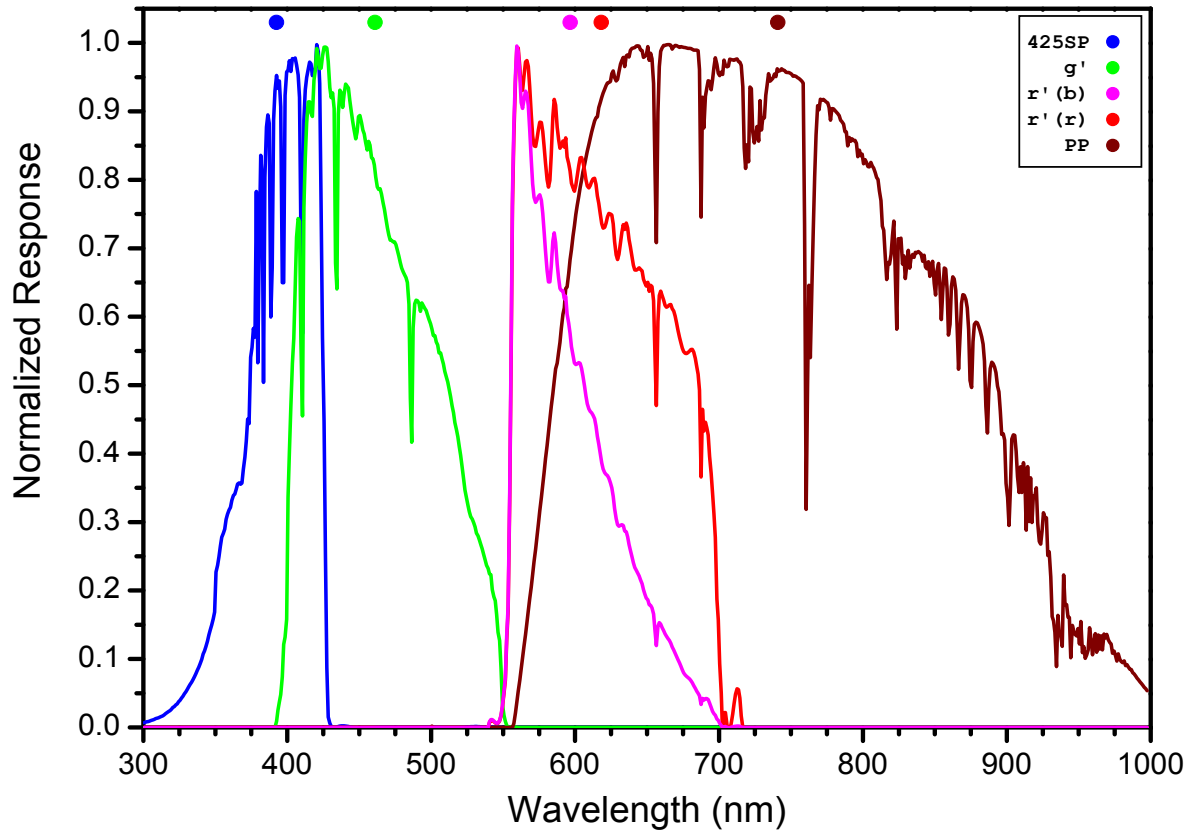

**Supplementary Figure 1 | HIPPI and PlanetPol effective bandpasses.**

The effective bandpasses used to observe Regulus. Colored dots, representing the effective wavelength of each band, are shown above a detailed normalized band response. The bandpass model begins with a model spectrum for Regulus with $\omega/\omega_c$ = 0.95 and *i* = 90 degrees, and modifies the spectrum to take account of filter, detector and modulator responses as well as absorption by the atmosphere. The atmosphere model has been tailored for the observing sites of the AAT and WHT. An airmass of 1.0 was assumed, but higher airmasses result in negligible differences. The bandpass model is calculated with 0.01 nm resolution but has been binned per nanometer here.

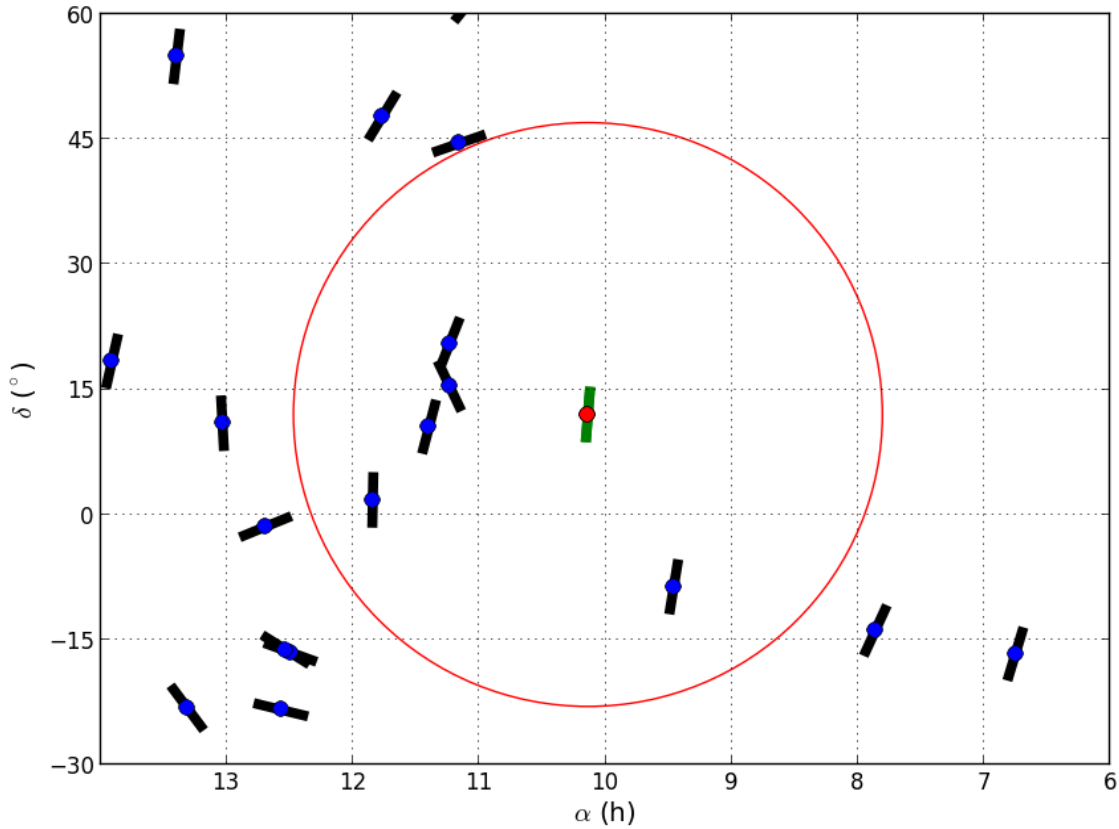

**Supplementary Figure 2 | The determined interstellar polarization position angle of Regulus from nearby stars.**

The red data point is Regulus, and the green bar its determined interstellar polarization position angle. The blue data points and black bars are nearby stars believed to be polarized only by the ISM and their position angles respectively. The plot shows the stars in right ascension (α) and declination (δ) with red circle denoting 35 degrees angular separation from Regulus, which represents the limit beyond which there is no correlation between the position angles of star pairs[28]. Position angle is defined in the range 0 – 180 degrees, and is measured anti-clockwise from 12 o'clock in the figure.

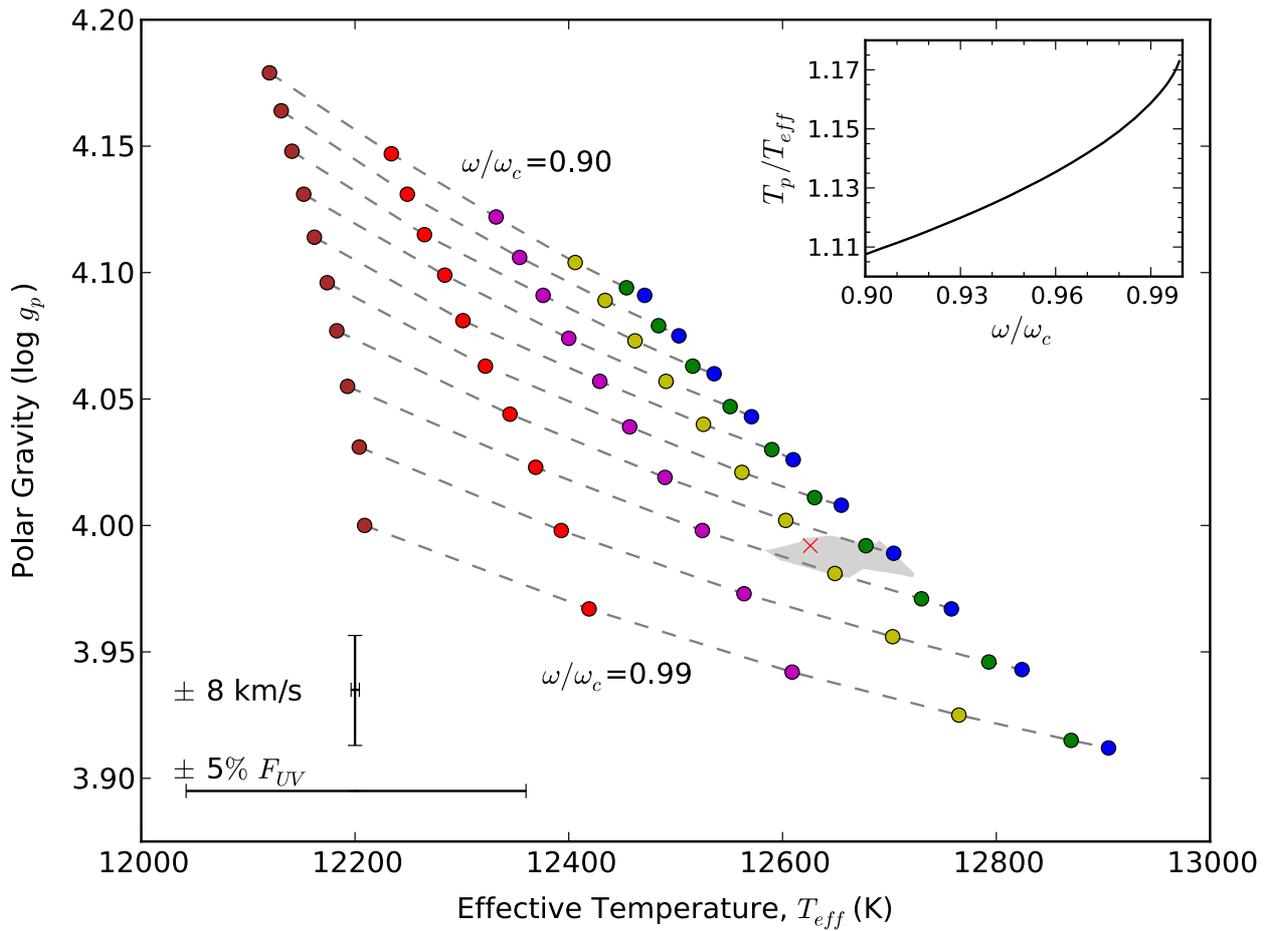

**Supplementary Figure 3 | The initial grid of log $g_p$ and $T_{eff}$ generated from assumed values of $\omega/\omega_c$ and *i*.**

Running from brown to blue across the plot the colored dots represent inclinations of 65 to 90 degrees in 5 degree increments. Each dashed line corresponds to a line of constant $\omega/\omega_c$ from 0.90 at the top of the plot to 0.99 at the bottom in increments of 0.01. Representative error bars associated with the measurement uncertainties in $F_{UV}$ and the *v* sin *i* determination are shown at the bottom left of the plot. A red cross is shown to represent the parameters belonging to our best-fit model, while the grey shaded region shows the 1-sigma modeling region based on chi-squared fitting transformed from Figure 4. Our radiative-transfer models take $T_p$ as the starting point rather than $T_{eff}$, and we show a conversion between the two in the inset.

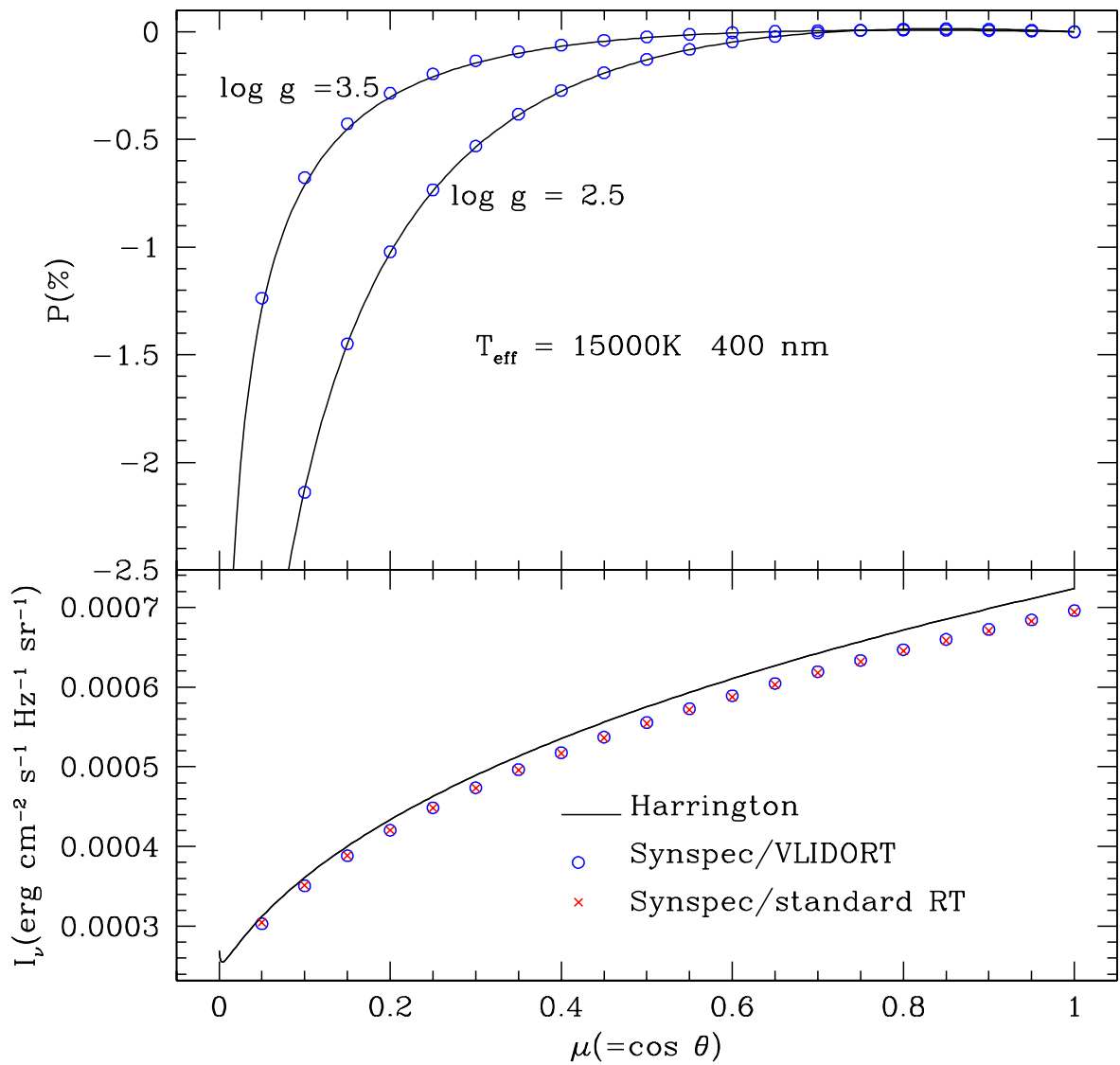

**Supplementary Figure 4 | Comparison of SYNSPEC/VILIDORT with the results of Harrington.**

Comparison of our SYNSPEC/VLIDORT radiative transfer with results from Harrington[17] for a TLUSTY model with $T_{eff}$ = 15000 K and two different gravities at a wavelength of 400 nm. The top panel shows Harrington's results (solid line) and the SYNSPEC/VLIDORT results (blue circles) for polarization as a function of $\mu$. The bottom panel shows specific intensities (at log $g$ = 3.5 only). The intensities from Harrington are a little higher than those from SYNSPEC, but the SYNSPEC/VLIDORT results agree very well with intensities from SYNSPEC's standard radiative transfer (red crosses).

Cotton, D. V. *et al.*, Polarization due to rotational distortion in the bright star Regulus. *Nature Astronomy* **1** (10), 690–696 (2017).

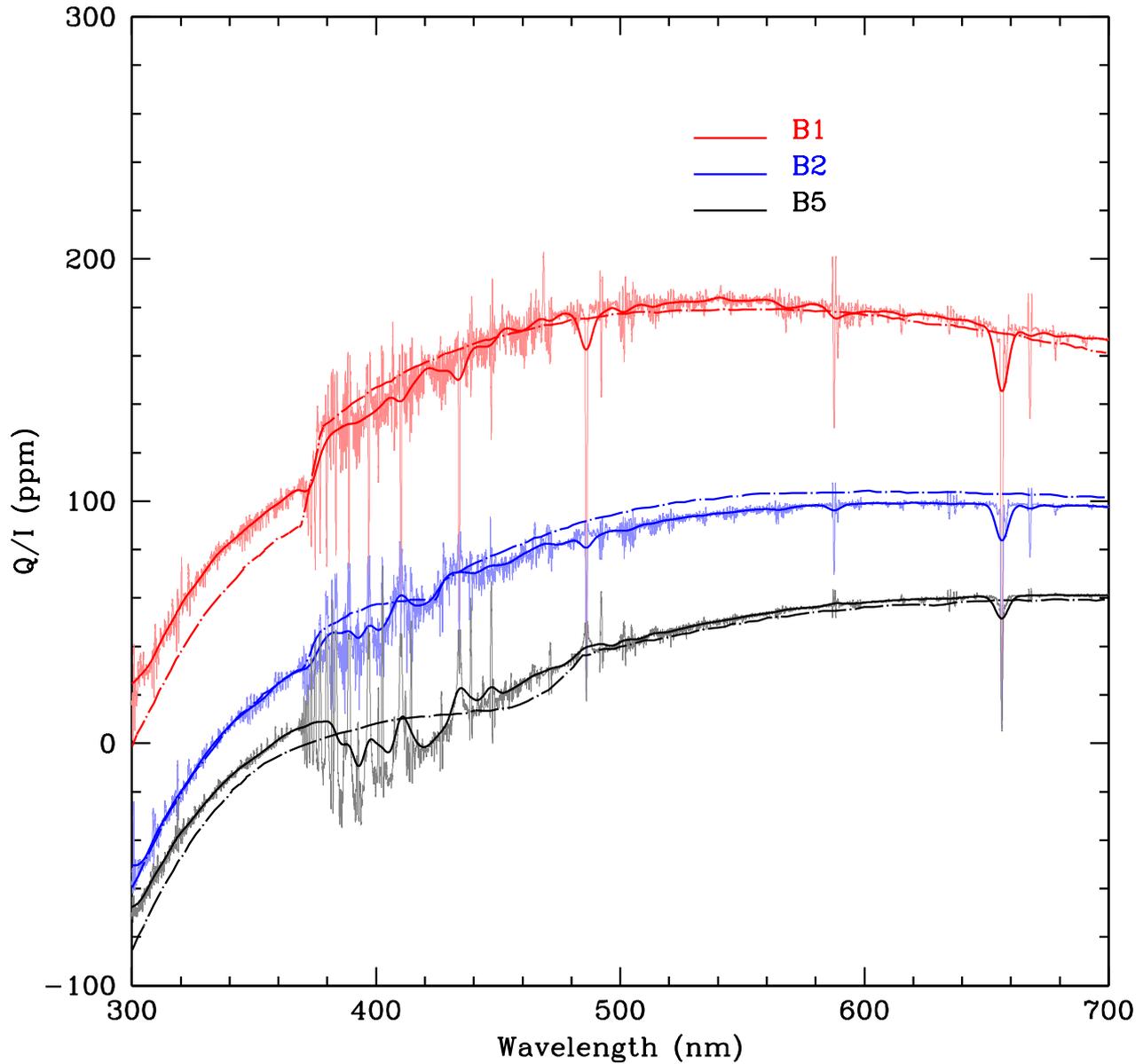

**Supplementary Figure 5 | Comparison of SYNSPEC/VILIDORT with the results of Sonneborn.**

Polarization for rapidly rotating B stars with $\omega/\omega_c$ = 0.95 and inclination 90 degrees as calculated by our modeling code (solid lines) compared with results by Sonneborn[8] (dot-dashed lines). Our modeling produces results at high spectral resolution shown by the light colored line here. The heavy line shows the smoothed data. Smoothed values like this are used in Figure 3, but the full resolution of the models is used as input to the bandpass model to facilitate the comparison with observation in Figure 4.

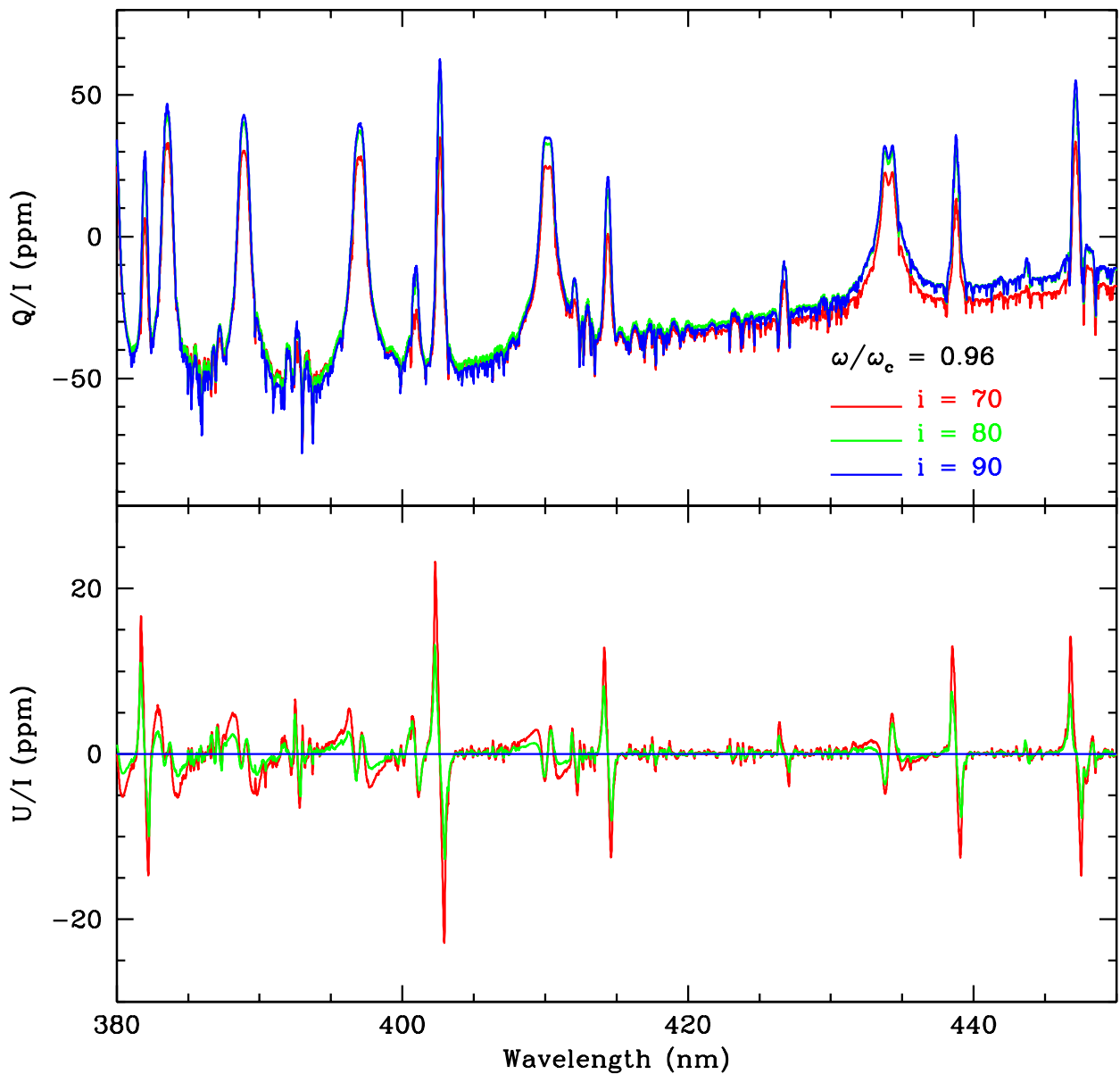

**Supplementary Figure 6 | A demonstration of the Öhman Effect in the range 380 to 450 nm for Regulus.**

The region of the spectrum from 380 to 450 nm corresponding to polarization models of Regulus for *ω/ω$_c$* = 0.96 and inclinations of 70, 80 and 90 degrees. Both Q/I and U/I are shown. It can be seen that the polarization across spectral lines is more sensitive to differences in inclination, at high inclinations, than is the broadband polarization shown in Figure 3.

**Supplementary Table 1 | Modeled effective wavelengths and polarization corrections by band.**

| Spec Type /Star | 425SP | | SDSS g' | | SDSS r' (b)* | | SDSS r' (r) | | PlanetPol** | |
|---|---|---|---|---|---|---|---|---|---|---|
| | λ (nm) | Eff. (%) | λ (nm) | Eff. (%) | λ (nm) | Eff. (%) | λ (nm) | Eff. (%) | λ (nm) | Eff. (%) |
| B0 | 387.5 | 45.2 | 459.1 | 87.7 | 595.7 | 85.3 | 616.8 | 81.4 | 735.0 | 91.5 |
| A0 | 397.1 | 54.7 | 462.2 | 88.6 | 596.6 | 85.1 | 618.3 | 81.1 | 745.4 | 92.3 |
| F0 | 396.3 | 53.9 | 466.2 | 89.6 | 598.3 | 84.8 | 620.8 | 80.7 | 753.8 | 93.0 |
| G0 | 395.0 | 52.5 | 470.7 | 90.6 | 599.9 | 84.5 | 623.0 | 80.2 | 760.5 | 93.5 |
| K0 | 396.8 | 54.2 | 474.4 | 91.6 | 601.0 | 84.3 | 624.5 | 80.0 | 765.6 | 93.9 |
| M0 | 399.6 | 57.0 | 477.5 | 92.0 | 604.6 | 83.7 | 629.3 | 79.1 | 791.4 | 95.1 |
| M5 | 399.5 | 56.9 | 477.3 | 91.7 | 605.4 | 83.5 | 630.4 | 78.9 | 804.4 | 95.6 |
| Regulus | 392.5 | 50.8 | 461.0 | 88.2 | 596.7 | 85.1 | 618.3 | 81.1 | 740.9 | 91.9 |

**Notes:** * Tabulated for B0 to M5 are the values associated with the BNS modulator, the efficiencies using the Micron Technologies modulator are systematically higher by ~1.5%.

** The B0 to M5 models shown for PlanetPol are those given by Hough et al.[13]. Whereas we linearly interpolate for intermediate spectral types observed with HIPPI, PlanetPol processing used the closest model. For Regulus we have recreated the PlanetPol bandpass model and run it with the Regulus stellar model. Note that this model goes to 1000 nm, whereas the PlanetPol bandpass extends beyond this to 1050 nm, albeit with very little sensitivity at the additional wavelengths.

**Supplementary Table 2 | High-polarization standards observed for position angle calibration.**

| Star (HD) | p (%) | θ (deg) | Ref | Observations per Run | | | |
|---|---|---|---|---|---|---|---|
| | | | | May 2014 | Feb 2016 | Jun 2016 | Dec 2016 |
| 84810 | 1.6 | 100.0 | 32,57 | 0 | 0 | 0 | 2 |
| 80558 | 3.3 | 161.8 | 32 | 0 | 1 | 0 | 1 |
| 147084 | 4.3 | 32.0 | 58,59 | 1 | 1 | 1 | 0 |
| 154445 | 3.4 | 90.1 | 32,57,59 | 2 | 0 | 1 | 0 |
| 160529 | 7.0 | 20.4 | 32,57 | 1 | 0 | 0 | 0 |
| 187929 | 1.8 | 93.8 | 32 | 1 | 0 | 0 | 0 |

**Supplementary Table 3 | Low-polarization standards and their assigned values in each band.**

| Star | Q/I (ppm) | U/I (ppm) |
|---|---|---|
| *425SP* | | |
| Sirius | 0.7 ± 0.6 | -1.9 ± 0.2 |
| β Hyi | -1.5 ± 1.1 | -5.7 ± 0.4 |
| β Vir | 3.4 ± 1.4 | -0.1 ± 1.4 |
| *g'* | | |
| Sirius | 0.7 ± 0.6 | -1.9 ± 0.2 |
| β Hyi | -1.5 ± 1.1 | -5.8 ± 0.4 |
| β Vir | 3.4 ± 1.4 | -0.1 ± 1.4 |
| *r' (b)* | | |
| Sirius | 0.7 ± 0.6 | -1.9 ± 0.2 |
| β Leo | 0.8 ± 1.1 | 2.3 ± 1.1 |
| α Ser | -2.4 ± 0.9 | 4.0 ± 1.0 |
| *r' (r)* | | |
| Sirius | 0.7 ± 0.6 | -1.9 ± 0.2 |
| β Vir | 3.4 ± 1.4 | -0.1 ± 1.4 |
| *PP\** | | |
| β Vir | 3.3 ± 1.4 | -0.1 ± 1.4 |
| β Leo | 0.8 ± 1.1 | 2.2 ± 1.1 |
| α Ser | -2.3 ± 0.9 | 3.9 ± 1.0 |

**Notes:** * Values for PlanetPol band represent the original measurements from Bailey et al.[15]; these were used only to derive the values in other bands, not for the corrections themselves.

In the case of those stars observed with PlanetPol, the formal 1-sigma errors are propagated from the internal measurement uncertainties (standard deviations), whilst for the other two the errors are defined by the uncertainty in the interstellar polarization with distance trend used to determine *Q/I* and *U/I*.

**Supplementary Table 4 | Observations of low-polarization standards for TP calibration.**

| Star | Band | Date | Time (UT) | Sky* (D/S) | Exp (s) | Q/I (ppm) | U/I (ppm) |
|---|---|---|---|---|---|---|---|
| β Leo | r' (b) | 2014-05-11 | 13:39:36 | S | 2560 | -22.4 ± 3.5 | -51.9 ± 3.1 |
| α Ser | r' (b) | 2014-05-11 | 15:49:02 | S | 2560 | -1.4 ± 2.7 | -59.7 ± 2.8 |
| β Leo | r' (b) | 2014-05-12 | 08:43:20 | S | 2560 | -31.7 ± 2.8 | -67.3 ± 2.7 |
| **May 2014 – r' (b)** | | | | | **TP:** | **-17.6 ± 12.7** | **-60.3 ± 5.2** |
| Sirius | 425SP | 2016-02-26 | 13:44:25 | S | 320 | -19.6 ± 2.9 | -1.2 ± 3.1 |
| Sirius | 425SP | 2016-02-28 | 09:48:04 | S | 320 | -26.4 ± 1.6 | -2.8 ± 1.6 |
| β Vir | 425SP | 2016-02-29 | 14:41:40 | S | 1280 | -11.6 ± 8.7 | 19.0 ± 10.0 |
| **Feb 2016 – 425SP** | | | | | **TP:** | **-24.5 ± 2.9** | **-2.0 ± 1.4** |
| Sirius | g' | 2016-02-26 | 09:55:36 | - | 320 | -18.9 ± 1.0 | 1.6 ± 1.0 |
| Sirius | g' | 2016-02-26 | 13:46:26 | S | 320 | -18.5 ± 1.9 | 4.4 ± 2.2 |
| β Vir | g' | 2016-02-26 | 17:37:01 | S | 640 | -21.0 ± 6.4 | 5.4 ± 7.1 |
| Sirius | g' | 2016-02-27 | 09:40:23 | S | 280 | -27.6 ± 3.9 | 6.3 ± 3.6 |
| Sirius | g' | 2016-02-28 | 09:31:20 | S | 320 | -17.1 ± 1.1 | 4.4 ± 1.1 |
| β Vir | g' | 2016-06-25 | 08:43:24 | S | 640 | -19.3 ± 9.1 | 0.2 ± 9.3 |
| β Vir | g' | 2016-06-25 | 09:09:42 | S | 640 | -25.2 ± 6.8 | 11.7 ± 7.6 |
| β Hyi | g' | 2016-06-25 | 14:18:17 | S | 640 | -13.9 ± 7.8 | -11.5 ± 10.3 |
| β Hyi | g' | 2016-06-25 | 16:42:10 | S | 640 | -24.6 ± 4.2 | 4.6 ± 4.4 |
| **Feb 2016 – g'** | | | | | **TP:** | **-18.7 ± 5.9** | **3.2 ± 4.5** |
| Sirius | r' (r) | 2016-02-27 | 13:27:31 | D | 320 | -10.6 ± 2.4 | -2.9 ± 2.3 |
| β Vir | r' (r) | 2016-02-27 | 16:08:20 | S | 1280 | -11.1 ± 4.8 | 8.3 ± 5.8 |
| Sirius | r' (r) | 2016-03-01 | 13:42:45 | S | 320 | -9.4 ± 1.4 | -3.8 ± 1.5 |
| **Feb 2016 – r' (r)** | | | | | **TP:** | **-9.8 ± 1.2** | **-3.0 ± 1.2** |
| Sirius | 425SP | 2016-11-30 | 17:29:39 | S | 320 | -29.3 ± 3.7 | 15.8 ± 3.9 |
| β Hyi | 425SP | 2016-12-01 | 10:17:47 | S | 640 | -28.2 ± 12.6 | 18.4 ± 12.8 |
| Sirius | 425SP | 2016-12-01 | 15:36:42 | D | 320 | -20.2 ± 2.5 | -1.5 ± 2.1 |
| Sirius | 425SP | 2016-12-07 | 17:35:21 | S | 320 | -28.8 ± 4.9 | 4.3 ± 8.2 |
| **Dec 2016 – 425SP** | | | | | **TP:** | **-23.9 ± 4.0** | **2.7 ± 4.3** |
| β Hyi | g' | 2016-11-30 | 11:08:40 | S | 640 | -30.0 ± 6.3 | -4.5 ± 6.2 |
| Sirius | g' | 2016-11-30 | 17:43:01 | S | 320 | -26.0 ± 1.8 | 1.3 ± 1.8 |
| Sirius | g' | 2016-12-02 | 15:39:03 | D | 320 | -24.1 ± 1.1 | 0.3 ± 1.2 |
| **Dec 2016 – g'** | | | | | **TP:** | **-24.8 ± 0.9** | **0.4 ± 1.0** |
| Sirius | r' (b) | 2016-12-04 | 14:00:32 | S | 640 | -23.2 ± 2.9 | 6.5 ± 3.4 |
| Sirius | r' (b) | 2016-12-05 | 15:57:56 | S | 640 | -15.2 ± 2.0 | -3.4 ± 2.2 |
| **Dec 2016 – r' (b)** | | | | | **TP:** | **-17.9 ± 3.6** | **-0.3 ± 4.2** |

**Notes:** * S indicates full sky subtraction, D indicates dark subtraction only, - indicates no subtraction applied.

The data have been bandpass corrected and rotated into the equatorial frame.

The stated errors in the individual observations are the internal standard deviations of the measurements. For the TP determination we use either the root-mean-square of the errors propagated in taking the mean of the individual measurements, or the error-weighted mean of the scatter (standard deviation) in the individual measurements, whichever is greater.

**Supplementary Table 5 | Observations of Regulus.**

| Date | Time (UT) | Exp (s) | Q/I (ppm) | U/I (ppm) | p (ppm) | θ (deg) |
|---|---|---|---|---|---|---|
| *Regulus – 425SP* | | | | | | |
| 2016-02-28 | 12:55:51 | 2560 | 31.7 ± 3.5 | -4.5 ± 2.5 | 32.0 ± 3.0 | 176 ± 2 |
| 2016-12-01 | 17:09:59 | 2560 | 22.4 ± 4.7 | -8.4 ± 4.9 | 23.9 ± 4.8 | 170 ± 6 |
| *Regulus – g'* | | | | | | |
| 2016-02-26 | 14:54:57 | 1280 | 12.1 ± 6.2 | 12.7 ± 5.0 | 17.5 ± 5.6 | 23 ± 9 |
| 2016-03-01 | 13:00:15 | 1600 | 8.8 ± 6.0 | 2.2 ± 4.7 | 9.1 ± 5.3 | 7 ±15 |
| 2016-12-02 | 16:28:40 | 960 | 11.3 ± 2.0 | -1.4 ± 1.9 | 11.4 ± 2.0 | 176 ± 5 |
| *Regulus – r' (b)* | | | | | | |
| 2014-05-11 | 08:27:17 | 2080 | -16.6 ±12.8 | 7.6 ± 5.7 | 18.3 ± 9.3 | 78 ±11 |
| 2016-12-04 | 14:00:32 | 1600 | -26.8 ± 8.3 | 14.7 ± 8.4 | 30.6 ± 8.3 | 76 ± 8 |
| *Regulus – r' (r)* | | | | | | |
| 2016-02-27 | 14:11:23 | 2560 | -27.3 ± 2.4 | 15.3 ± 2.4 | 31.3 ± 2.4 | 75 ± 2 |
| 2016-03-01 | 14:42:24 | 2560 | -25.3 ± 1.7 | 14.1 ± 1.8 | 28.9 ± 1.8 | 76 ± 2 |
| *Regulus – PP* | | | | | | |
| 2005-04-26 | 21:16:53 | 1440 | -34.8 ± 1.1 | 12.5 ± 1.0 | 37.0 ± 1.0 | 80 ± 1 |
| 2006-02-20 | 04:59:35 | 1440 | -33.4 ± 1.1 | 15.2 ± 1.1 | 36.7 ± 1.1 | 78 ± 1 |
| *HD 99028 – g'* | | | | | | |
| 2016-02-26 | 14:54:57 | 1280 | 6.8 ± 5.4 | -3.8 ± 5.4 | 7.8 ± 5.4 | 166 ±20 |

**Notes:** Values are post application of the bandpass model; position angle calibration and TP subtraction and the uncertainties include the errors associated with those operations. Also included are observations of the interstellar control star HD 99028.

The errors in *Q/I* and *U/I* are the root-mean-square errors propagated from the error-weighted mean of the individual observations. The errors in *p* and *θ* are propagated from those in *Q/I* and *U/I*.

**Supplementary Table 6 | Control stars used for interstellar polarization angle determination, and the adopted interstellar polarization for Regulus in the g' band.**

| Star (HD) | Pos (RA, Dec) | Band | d (pc) | p (ppm)* | θ (deg) | Wt | Ref |
|---|---|---|---|---|---|---|---|
| 81797 | 09.46h, -08.7° | g' | 55.3 | 8.8±3.7 | 171±12 | 0.34 | 26 |
| 97603 | 11.24h, +20.5° | PP | 17.9 | 3.7±2.4 | 159±23 | 0.49 | 15 |
| 97633 | 11.24h, +15.4° | PP | 50.6 | 6.9±2.7 | 25±13 | 0.53 | 15 |
| 99028 | 11.40h, +10.5° | g' | 23.7 | 7.9±5.4 | 166±20 | 0.46 | New |
| 102870 | 11.84h, +01.8° | PP | 10.9 | 3.3±1.4 | 179±10 | 0.22 | 15 |
| **Regulus** | **10.14h, +12.0°** | **g'** | **24.3** | **6.3±1.4** | **175±12** | | |

**Notes:** * The *p* given here for the control stars is an undebiased value; and the error is the internal standard deviation of the measurement, propagated from *Q/I* and *U/I* To calculate *p/d* and thus the scatter used as the error in the determined interstellar polarization of Regulus, *p* is debiased as $\hat{p} = \sqrt{p^2 - \sigma_p^2}$. See the text for the method of calculating interstellar *p* for Regulus.

**Supplementary Table 7 | Combined Regulus measurements before and after interstellar subtraction, and after rotation corresponding to the determined position angle.**

| Band | Q/I | U/I | p | θ |
|---|---|---|---|---|
| *Error Weighted Observations* | | | | |
| 425SP | 28.4 ± 2.8 | -5.3 ± 2.2 | 28.8 ± 2.5 | 175 ± 2 |
| g' | 11.1 ± 1.8 | 0.7 ± 1.7 | 11.2 ± 1.7 | 2 ± 4 |
| r' (b) | -23.8 ± 7.0 | 9.8 ± 4.7 | 25.8 ± 5.8 | 79 ± 6 |
| r' (r) | -26.0 ± 1.4 | 14.5 ± 1.4 | 29.7 ± 1.4 | 75 ± 1 |
| PP | -34.1 ± 0.8 | 13.7 ± 0.7 | 36.8 ± 0.8 | 79 ± 1 |
| *After Interstellar Subtraction* | | | | |
| 425SP | 22.2 ± 3.1 | -4.3 ± 2.3 | 22.7 ± 2.7 | 175 ± 3 |
| g' | 4.9 ± 2.3 | 1.7 ± 1.8 | 5.2 ± 2.0 | 10 ± 10 |
| r' (b) | -29.8 ± 7.1 | 10.8 ± 4.7 | 31.7 ± 5.9 | 80 ± 5 |
| r' (r) | -31.9 ± 1.9 | 15.4 ± 1.5 | 35.4 ± 1.7 | 77 ± 1 |
| PP | -39.5 ± 1.5 | 14.6 ± 0.9 | 42.1 ± 1.2 | 80 ± 1 |
| *After Rotation into the Axial Frame* | | | | |
| Band | $Q_r/I$ | $U_r/I$ | p | $θ_r$ |
| 425SP | -22.3 ± 2.4 | -3.9 ± 3.0 | 22.7 ± 2.7 | 95 ± 4 |
| g' | -4.0 ± 1.9 | -3.3 ± 2.2 | 5.2 ± 1.4 | 110 ± 11 |
| r' (b) | 31.7 ± 5.2 | 0.5 ± 6.7 | 31.7 ± 6.0 | 0 ± 6 |
| r' (r) | 35.3 ± 1.6 | -3.0 ± 1.9 | 35.4 ± 1.8 | 178 ± 2 |
| PP | 42.1 ± 1.1 | 0.5 ± 1.4 | 42.1 ± 1.3 | 0 ± 1 |

**Notes:** The errors in *Q/I* and *U/I* are the root-mean-square errors propagated from the error-weighted mean of the individual observations; those after interstellar subtraction incorporate the error in the determined interstellar polarization as a root-mean-square addition. The errors in *p* and *θ* are propagated from those in *Q/I* and *U/I*.

**Supplementary Table 8 | Adopted and final parameters for Regulus.**

| Adopted Stellar Parameters | | Notes/References |
|---|---|---|
| $v \sin i$ (km/s) | 318 ± 8 | a, [37-40] |
| V (mag) | 1.35 ± 0.02 | [60] |
| d (pc) | 24.3 ± 0.2 | [31] |
| $M_V$ (mag) | -0.58 | a |
| | | |
| *Interstellar Medium* | | |
| $\lambda_{max}$ (nm) | 470 | [27] |
| $p_{max}$ (ppm) | 6.3 ± 1.4 | a, [28] |
| $\theta_{ISM}$ (deg) | 175 ± 12 | b, [28] |
| | | |
| *Polarimetric Determination* | | |
| $\theta_R$ (deg) | 79.5 ± 0.7 | c |
| | | |
| *Modeling Determination* | | |
| i (deg) | 80.0 (> 76.5) | c |
| $\omega/\omega_c$ | $0.965^{+0.006}_{-0.008}$ | c |
| log $g_p$ | $3.992^{+0.022}_{-0.026}$ | c |
| log $g_{eq}$ | $3.333^{+0.095}_{-0.099}$ | c |
| $T_p$ (K) | $14375^{+215}_{-181}$ | c |
| $T_{eq}$ (K) | $11060^{+138}_{-140}$ | c |
| $T_{eff}$ (K) | $12626^{+190}_{-163}$ | c |
| | | |
| *Other Derived Physical Parameters* | | |
| Mass ($M_{Sun}$) | $3.72^{+0.56}_{-0.47}$ | d |
| $r_p$ ($r_{Sun}$) | $3.22^{+0.29}_{-0.25}$ | d |
| $r_{eq}$ ($r_{Sun}$) | $4.22^{+0.35}_{-0.31}$ | d |
| $P_{rot}$ (d) | $0.662^{+0.046}_{-0.042}$ | d |
| log ($L/L_{Sun}$) | $2.520^{+0.076}_{-0.075}$ | d |
| $v_e/v_c$ | $0.844^{+0.015}_{-0.017}$ | d |

**Notes:** (a) calculated based on available data, (b) one additional measurement was added to those available in the literature to calculate this parameter, (c) determined from measurements and models first presented in this work, (d) calculated based on the determined and adopted parameters.

See the Methods section of the text for a description of how the errors were determined for each quantity. Specifically, for data not taken directly from the literature, this information can be found in the sections headed Interstellar polarization subtraction, Determination of Regulus' rotation-axis position angle, Determination of projected equatorial rotational velocity, Constraining stellar parameter space, and Regulus final parameters.

Cotton, D. V. *et al.*, Polarization due to rotational distortion in the bright star Regulus. *Nature Astronomy* **1** (10), 690–696 (2017).

**Note added to arXiv version:** This article is the subject of a Nature Astronomy 'news & views' piece that appeared in the same issue of the journal:

Harrington, J. Patrick. Polarization from a spinning star, *Nature Astronomy* **1** (10), 657–658 (2017).